# Exploring the Adoption Intention in Using AI-Enabled Educational Tools Among Preservice Teachers in the Philippines: A Partial-Least Square Modeling

Vanessa Baltazar Sibug
College of Education
Pampanga State University
Bacolor, Pampanga, Philippines
vbsibug@pampangastateu.edu.ph

Emerson Quiambao Fernando
College of Education
Pampanga State University
Lubao, Pampanga, Philippines
eqfernando@pampangastateu.edu.ph

Almer Balingit Gamboa
College of Education
Pampanga State University
Mexico, Pampanga, Philippines
abgamboa@pampangastateu.edu.ph

Roque Francis Badajos Dianelo
College of Education
Pampanga State University
Mexico, Pampanga, Philippines
rfbdianelo@pampangastateu.edu.ph

Agnes Romero Regala
College of Education
Pampanga State University
Bacolor, Pampanga, Philippines
arregala@pampangastateu.edu.ph

Joseph Alexander Bansil
College of Education
Pampanga State University
Lubao, Pampanga, Philippines
jabansil@pampangastateu.edu.ph

Jan Henry Beltran Sunga
College of Education
Pampanga State University
Bacolor, Pampanga, Philippines
jhbsunga@pampangastateu.edu.ph

Vernon Grace Magat Maniago
College of Computing Studies
Pampanga State University
Sto. Tomas, Pampanga, Philippines
vgmmaniago@pampangastateu.edu.ph

John Paul Palo Miranda*
College of Computing Studies
Pampanga State University
Mexico, Pampanga, Philippines
jppmiranda@pampangastateu.edu.ph

## Abstract
This study examines the factors influencing pre-service teachers' behavioral intention to use AI-enabled educational tools during their practicum, using the Unified Theory of Acceptance and Use of Technology 2 (UTAUT2) as the theoretical framework. The model includes the core UTAUT2 constructs such as performance expectancy, effort expectancy, hedonic motivation, social influence, facilitating conditions, price value, and habit. It also incorporates additional predictors including computer self-efficacy, computer anxiety, and computer playfulness. Data were collected from 563 pre-service teachers using a structured questionnaire and analyzed using Partial Least Squares Structural Equation Modeling (PLS-SEM). The results indicate that performance expectancy and hedonic motivation are the strongest predictors of behavioral intention. Computer self-efficacy, computer anxiety, and computer playfulness significantly influenced effort expectancy, although effort expectancy did not directly predict behavioral intention. Performance expectancy was significantly predicted by extrinsic motivation, job fit, relative advantage, and outcome expectations. Constructs such as social influence and facilitating conditions showed limited or inverse effects. These findings suggest that internal motivational, cognitive, and emotional factors are more influential than external or institutional factors in shaping the adoption of AI-enabled tools. The study highlights the importance of promoting personal relevance, confidence, and enjoyment in teacher preparation programs to encourage technology integration.

## 1 Introduction

Educational technology tools (ETTs) refer to digital applications that support and enhance teaching and learning processes in both traditional and non-traditional educational settings [4, 6, 13, 16]. These tools include learning management systems, video conferencing platforms, mobile learning applications, virtual reality environments, and artificial intelligence (AI)-enabled systems designed to assist with instruction and assessment tasks [21]. Although ETTs are increasingly integrated into teacher education and classroom practice, their effective use depends not only on access or infrastructure but also on how future educators perceive, adopt, and apply these tools in their professional development [7].

Understanding how pre-service teachers adopt educational technology is particularly important during the practicum phase. This period marks the early formation of teaching practices and pedagogical approaches. Previous studies have shown that teachers' beliefs, attitudes, and emotional responses toward technology influence their willingness to use it. These responses include confidence in using digital tools, anxiety or apprehension about new technologies, and the perceived instructional value of such tools [8]. However, there remains a significant gap in understanding how pre-service teachers navigate the adoption of AI-enabled educational tools during the practicum. This gap limits the ability of teacher education programs to prepare future educators for AI-integrated classrooms. Furthermore, identifying and addressing both personal and contextual influences is seen as essential for promoting meaningful and sustained integration of technology in teacher preparation programs [5].

To examine these influences, the present study adopts the Unified Theory of Acceptance and Use of Technology 2 (UTAUT2) developed by Venkatesh et al. [11, 20]. This theoretical model explains technology acceptance through several constructs, including performance expectancy, effort expectancy, social influence, facilitating conditions, hedonic motivation, price value, and habit [11, 20]. UTAUT2 extends the original UTAUT framework by incorporating both utilitarian and hedonic motivations. It has been successfully applied in diverse educational contexts, such as mobile learning, augmented reality, massive open online courses (MOOCs), and blended learning environments [2, 3, 14].

In this study, ETT refers specifically to AI-enabled tools that pre-service teachers may use during their practicum to support instructional planning, classroom engagement, assessment, and teaching strategies. The proposed structural model incorporates additional constructs, including computer self-efficacy, computer anxiety, and computer playfulness, as antecedents of effort expectancy and hedonic motivation. This model allows the examination of both direct and indirect effects among the variables that influence behavioral intention. By focusing on pre-service teachers during the practicum, this study contributes to a deeper understanding of the factors that motivate or inhibit the adoption of AI-enabled educational tools. The results aim to inform teacher education programs and support the development of strategies that align with pre-service teachers' readiness, motivation, and professional needs.

## 2 Materials and Methods

### 2.1 Study Design, Participants, and Recruitment

The study utilized a descriptive cross-sectional research design to assess the intention to adopt AI-enabled educational tools among preservice teachers during their practicum in Central Luzon, Philippines. Participants were recruited through purposive sampling targeting teacher education institutions in the region. The inclusion criteria required participants to be currently enrolled in a practicum course during the data collection period. Data were gathered using an online survey administered between October 4, 2024, and March 31, 2025. Informed consent was obtained electronically before participation.

### 2.2 Survey Instrument, Data Collection, and Analysis

The study used a structured questionnaire based on the UTAUT2 to measure the intention to adopt ETT among preservice teachers. The questionnaire included items adapted from validated instruments used in previous studies on technology adoption. A four-point Likert scale measured the level of agreement for each item, ranging from "strongly disagree" to "strongly agree." The first section of the instrument collected basic demographic information.

To ensure content validity, a panel of three educational technology professors reviewed the instrument. For reliability testing, 50 preservice teachers completed a pilot version of the survey. Internal consistency was assessed using Cronbach's alpha, with a threshold of 0.70 as the minimum acceptable value. Revisions were made based on expert feedback and pilot results to improve item clarity and measurement reliability. For data collection, they were collected through an online survey administered via Google Forms.

In this study, AI-enabled ETT was contextualized to multiple platforms such as but not limited to software that assist in instructional tasks through AI. During their practicum, preservice teachers commonly used tools such as Canva, Slidesgo, and Quillbot for content creation; ChatGPT, Gemini, and Copilot for lesson planning and instructional support; Quizizz, Kahoot!, and Classcraft for student engagement and formative assessment; and Gradescope for automated grading and feedback.

The final dataset was analyzed using partial least squares structural equation modeling (PLS-SEM) to examine the relationships among variables and test the proposed framework. This approach allowed for the identification of key predictors and the evaluation of the model's explanatory power in the context of ETT adoption.

### 2.3 Study Hypotheses

The proposed model investigates the determinants of pre-service teachers' behavioral intention to use AI-enabled educational tools. Based on the UTAUT2 and related technology adoption frameworks, the study incorporates individual, motivational, and contextual factors as predictors of behavioral intention. Computer Self-Efficacy (CSE) refers to the belief in one's ability to use technology effectively. It is expected to positively influence perceived ease of use and thereby effort expectancy. Conversely, Computer Anxiety (CA), defined as discomfort or fear associated with using technology, is anticipated to negatively influence effort expectancy. Computer Playfulness (CP), which reflects a user's spontaneous interaction with technology, is expected to positively affect both effort expectancy and hedonic motivation. Effort Expectancy (EE), or the perceived ease of using AI tools, is hypothesized to positively influence Behavioral Intention to Use (BITU). Similarly, Hedonic Motivation (HM), which reflects the perceived enjoyment derived from using technology, is also expected to enhance behavioral intention. Performance Expectancy (PE), referring to the belief that using AI tools will improve teaching or learning performance, is hypothesized to have a direct positive influence on behavioral intention. PE is further conceptualized as being shaped by Extrinsic

Motivation (EM), Job Fit (JF), Relative Advantage (RA), and Outcome Expectations (OE). In addition to individual-level factors, several contextual constructs are expected to predict behavioral intention. Facilitating Conditions (FC), such as available technical and institutional support, are hypothesized to positively influence adoption. Social Influence (SI), referring to the degree to which peers and supervisors encourage technology use, is also expected to affect behavioral intention. Price Value (PV), or the trade-off between perceived benefits and the cost of using AI tools, is expected to contribute positively. Finally, Habit (HB), which reflects prior routine use or exposure to technology, is hypothesized to predict future intention to use AI tools.

Based on the conceptual model, the following hypotheses are proposed. H1 posits that Computer Self-Efficacy (CSE) positively influences EE, while H2 suggests that CA negatively influences EE. H3 states that CP positively influences EE, and H4 proposes that CP also positively influences HM. H5 hypothesizes that EE positively influences BITU, and H6 states that HM positively influences BITU. H7 suggests that PE positively influences BITU. H8 posits that EM, JF, RA, and OE positively influence PE. H9 proposes that FC positively influence BITU, while H10 suggests that SI positively influences BITU. H11 states that PV positively influences BITU, and H12 hypothesizes that HB positively influences BITU.

## 3 Results and Discussion

### 3.1 Demographic Characteristics

A total of 563 pre-service teachers participated in the study, with a mean age of 21.06 years (SD = 2.13). Most respondents were female (71.40%, n = 402), and the rest were male (28.60%, n = 161). In terms of specialization, the majority were enrolled in Physical Education (50.09%, n = 282), followed by Elementary Education (24.51%, n = 138), Secondary Education (25.39%, n = 143) with majors in Social Studies, English, Technology and Livelihood Education, and Filipino. Regarding practicum deployment, 41.92% were assigned to Junior High School (n = 236), 29.48% to Tertiary or College level (n = 166), and 22.20% to Primary Education (n = 125), with others deployed to Senior High School (5.15%, n = 29), Kindergarten (1.24%, n = 7). In terms of educational technology experience, 38.90% had none (n = 219), 18.12% had 1–2 years (n = 102), 15.99% had less than 1 year (n = 90), 13.85% had more than 4 years (n = 78), and 13.21% had 3–4 years (n = 74). A total of 55.24% respondents reported no previous teaching experience (n = 311), while 44.76% reported having teaching experience (n = 252). Most were enrolled in public institutions (98%, n = 552), with only a few from private institutions (2%, n = 11). Regular access to technology was reported by 73.89% respondents (n = 416), while 23.80% had occasional access (n = 134), and 2.31% reported rarely or never having access (n = 13). For AI-related training, 39.61% were self-taught without formal training (n = 223), 22.91% had informal training (n = 129), 2.07% had formal training (n = 113), and 17.41% had no training (n = 98).

### 3.2 Reliability and Validity of the Measurement Model

All items showed strong and statistically significant factor loadings on their respective constructs (Table 1). Each loading met the acceptable threshold of .70, with values ranging from .819 to .972. The t-statistics from 1,000 bootstrap samples ranged from 33.499 to 611.143. All items had p-values less than .001. These values confirm that each item reflects its construct accurately and supports the convergent validity of the measurement model.

The Fornell–Larcker criterion was used to assess discriminant validity. The square root of the average variance extracted (AVE) was placed along the diagonal of the correlation matrix (Table 2). Each construct's square root AVE exceeded its correlation with any other construct in the model. These results confirm that all constructs achieved acceptable discriminant validity.

### 3.3 Hypothesis Testing

The structural model examined both direct and indirect effects and yielded several supported hypotheses (Table 3). For predictors of EE, significant positive effects were found from CSE ($\beta$ = .469, p < .001), CA ($\beta$ = .059, p = .007), and CP ($\beta$ = .387, p < .001), supporting H1, H2, and H3. CP also had a significant positive influence on HM ($\beta$ = .840, p < .001), supporting H4. Among the direct predictors of BITU, significant positive effects were observed from HM ($\beta$ = .292, p < .001) and PE ($\beta$ = .302, p < .001, supporting H6 and H7, while EE did not show a significant effect on BITU ($\beta$ = −.013, p = .830), leading to the rejection of H5. For the predictors of PE, all four were significant: EM ($\beta$ = .487, p < .001), JF ($\beta$ = .162, p < .001), RA ($\beta$ = .071, p = .033), and OE ($\beta$ = .236, p < .001), supporting H8a through H8d. FC did not significantly influence BITU ($\beta$ = −.029, p = .599), providing no support for H9. SI showed a significant negative effect ($\beta$ = −.096, p = .014), which was contrary to the expected direction, thus providing partial support for H1. PV ($\beta$ = .096, p = .078) and HB ($\beta$ = .081, p = .095) had marginal effects, which were considered partial support for H11 and H12. Overall, the model revealed strong support for motivational and expectancy-related predictors, with mixed results for contextual and habitual influences.

## 4 Discussion

The results of the PLS-SEM analysis align with the UTAUT and its extended version UTAUT2, particularly in identifying motivational and expectancy-related factors as primary drivers of behavioral intention to use AI-enabled educational tools among pre-service teachers. PE and HM emerged as significant predictors of BITU, which is consistent with UTAUT2's emphasis on both utilitarian value and enjoyment in influencing adoption decisions. The strong influence of PE confirms that when teachers perceive AI tools as beneficial to their instructional performance, they are more likely to adopt them. Similarly, the positive effect of HM suggests that intrinsic enjoyment significantly motivates future use of AI-enabled technologies. This result supports [18], who found that both performance expectancy and hedonic motivation significantly influenced pre-service teachers' behavioral intention to integrate GenAI tools in instruction. PE was also significantly influenced by EM, JF, RA, and OE, supporting the idea that perceived usefulness is shaped by the degree to which AI tools align with task goals, improve work relevance, and offer functional benefits [1]. Several studies have also confirmed that outcome expectations, job alignment, and perceived advantages enhanced educators' evaluations of usefulness [12, 17].

Table 1: List of scale items for adoption intention of AI-enabled tools for preservice teachers

| Construct | Item | FL | T-Statistic | p value | α | CR | AVE |
|---|---|---|---|---|---|---|---|
| BITU | BITU1 | .819 | 33.499 | < .001 | .835 | .903 | .756 |
|  | BITU2 | .894 | 548.904 | < .001 |  |  |  |
|  | BITU3 | .893 | 553.031 | < .001 |  |  |  |
| PE | PE1 | .937 | 596.218 | < .001 | .933 | .958 | .883 |
|  | PE2 | .948 | 569.38 | < .001 |  |  |  |
|  | PE3 | .933 | 561.573 | < .001 |  |  |  |
| EM | EM1 | .966 | 575.161 | < .001 | .930 | .966 | .935 |
|  | EM2 | .968 | 565.995 | < .001 |  |  |  |
| JF | JF1 | .967 | 55.956 | < .001 | .929 | .966 | .934 |
|  | JF2 | .966 | 563.771 | < .001 |  |  |  |
| RA | RA1 | .951 | 602.865 | < .001 | .895 | .950 | .905 |
|  | RA2 | .952 | 581.121 | < .001 |  |  |  |
| OE | OE1 | .964 | 6.180 | < .001 | .918 | .961 | .925 |
|  | OE2 | .960 | 584.788 | < .001 |  |  |  |
| HB | HB1 | .960 | 581.121 | < .001 | .913 | .959 | .920 |
|  | HB2 | .959 | 575.161 | < .001 |  |  |  |
| SI | SI1 | .972 | 533.274 | < .001 | .938 | .970 | .942 |
|  | SI2 | .970 | 582.336 | < .001 |  |  |  |
| PV | PV1 | .954 | 583.558 | < .001 | .901 | .953 | .910 |
|  | PV2 | .954 | 564.879 | < .001 |  |  |  |
| FC | FC1 | .955 | 577.523 | < .001 | .901 | .953 | .910 |
|  | FC2 | .953 | 548.904 | < .001 |  |  |  |
| HM | HM1 | .957 | 545.869 | < .001 | .905 | .955 | .914 |
|  | HM2 | .954 | 565.995 | < .001 |  |  |  |
| EE | EE1 | .964 | 578.715 | < .001 | .919 | .961 | .926 |
|  | EE2 | .960 | 587.271 | < .001 |  |  |  |
| CSE | CSE1 | .931 | 611.143 | < .001 | .856 | .933 | .875 |
|  | CSE2 | .940 | 592.334 | < .001 |  |  |  |
| CA | CA1 | .951 | 559.4 | < .001 | .896 | .950 | .905 |
|  | CA2 | .952 | 571.672 | < .001 |  |  |  |
| CP | CP1 | .970 | 569.38 | < .001 | .937 | .969 | .941 |
|  | CP2 | .970 | 567.116 | < .001 |  |  |  |

Table 2: Inter-construct correlations with square root of AVE

|  | BITU | PE | EM | JF | RA | OE | HB | SI | PV | FC | HM | EE | CSE | CA | CP |
|---|---|---|---|---|---|---|---|---|---|---|---|---|---|---|---|
| BITU | **.87** |  |  |  |  |  |  |  |  |  |  |  |  |  |  |
| PE | .64 | **.94** |  |  |  |  |  |  |  |  |  |  |  |  |  |
| EM | .64 | .87 | **.97** |  |  |  |  |  |  |  |  |  |  |  |  |
| JF | .61 | .82 | .83 | **.97** |  |  |  |  |  |  |  |  |  |  |  |
| RA | .56 | .77 | .77 | .80 | **.95** |  |  |  |  |  |  |  |  |  |  |
| OE | .60 | .81 | .79 | .81 | .80 | **.96** |  |  |  |  |  |  |  |  |  |
| HB | .58 | .73 | .74 | .80 | .74 | .78 | **.96** |  |  |  |  |  |  |  |  |
| SI | .46 | .65 | .66 | .69 | .65 | .65 | .71 | **.97** |  |  |  |  |  |  |  |
| PV | .57 | .73 | .74 | .75 | .71 | .77 | .77 | .79 | **.95** |  |  |  |  |  |  |
| FC | .56 | .74 | .75 | .75 | .73 | .77 | .74 | .70 | .78 | **.95** |  |  |  |  |  |
| HM | .63 | .75 | .77 | .77 | .74 | .77 | .79 | .72 | .80 | .83 | **.96** |  |  |  |  |
| EE | .57 | .73 | .75 | .74 | .71 | .75 | .74 | .67 | .78 | .81 | .85 | **.96** |  |  |  |
| CSE | .58 | .70 | .71 | .70 | .67 | .74 | .70 | .67 | .75 | .79 | .80 | .81 | **.94** |  |  |
| CA | .33 | .31 | .34 | .34 | .36 | .33 | .39 | .37 | .41 | .37 | .35 | .38 | .35 | **.95** |  |

Table 3: Hypothesis testing results

| # | Hypothesis | $\beta$ | p-value | Support |
|---|---|---|---|---|
| H1 | CSE → EE | .469 | .000 | Yes |
| H2 | CA → EE | .059 | .007 | Yes |
| H3 | CP → EE | .387 | .000 | Yes |
| H4 | CP → HM | .840 | .000 | Yes |
| H5 | EE → BITU | −.013 | .830 | No |
| H6 | HM → BITU | .292 | .000 | Yes |
| H7 | PE → BITU | .302 | .000 | Yes |
| H8a | EM → PE | .487 | .000 | Yes |
| H8b | JF → PE | .162 | .000 | Yes |
| H8c | RA → PE | .071 | .033 | Yes |
| H8d | OE → PE | .236 | .000 | Yes |
| H9 | FC → BITU | −.029 | .599 | No |
| H10 | SI → BITU | −.096 | .014 | Partial (negative) |
| H11 | PV → BITU | .096 | .078 | Partial (marginal) |
| H12 | HB → BITU | .081 | .095 | Partial (marginal) |

Although EE did not show a significant direct effect on behavioral intention, the construct was significantly shaped by CSE, CA, and CP. These findings are consistent with prior research that positions EE as a secondary construct influenced by cognitive and emotional traits rather than a direct driver of intention. [18] observed that GenAI anxiety negatively affected effort expectancy, while technology self-efficacy positively influenced it, supporting this indirect pathway. The pathway from CP to HM further confirmed that teachers who find digital tools playful are more likely to experience enjoyment when using them. This reinforces the role of user engagement in adoption and aligns with Venkatesh et al., who emphasized the influence of playfulness and intrinsic motivation in shaping technology acceptance. These results highlight the need to strengthen cognitive confidence and reduce anxiety to enhance perceptions of ease of use.

Unexpectedly, SI demonstrated a significant negative effect on behavioral intention. This suggests that external pressure or expectations from peers or institutions may not always promote adoption and, in some cases, may even reduce willingness to engage, particularly when such influence is perceived as misaligned with users' autonomy or readiness. Studies also found that excessive or misaligned social pressure can undermine individual motivation [9, 10, 15]. Furthermore, FC did not significantly influence behavioral intention, indicating that institutional support alone may not be sufficient to drive adoption in the absence of strong motivational or efficacy beliefs. [18] similarly reported that FC had no significant association with intention among pre-service teachers [19]. PV and HB showed marginal effects which implies that pre-service teachers may not consider cost-benefit judgments or previous behaviors central in their adoption decisions. Overall, these findings confirm UTAUT2's key assumptions while also pointing to important contextual and psychological pathways that training programs should consider when promoting AI adoption.

## 5 Conclusion and Recommendations

This study examined the determinants of pre-service teachers' behavioral intention to use AI-enabled educational tools during their practicum using the PLS-SEM approach, guided by the UTAUT and UTAUT2 frameworks. The results confirmed that motivational, expectancy-based, and cognitive factors significantly influenced adoption intention. Performance Expectancy and Hedonic Motivation were the strongest predictors of behavioral intention, emphasizing that perceived usefulness and enjoyment play critical roles in shaping technology adoption. Although Effort Expectancy did not have a direct effect, it was significantly influenced by Computer Self-Efficacy, Computer Anxiety, and Computer Playfulness, indicating that ease of use is driven by individual confidence and engagement. Performance Expectancy was also positively shaped by Extrinsic Motivation, Job Fit, Relative Advantage, and Outcome Expectations, highlighting the importance of aligning AI tools with personal and instructional relevance.

In contrast, contextual and external constructs such as Facilitating Conditions, Social Influence, Price Value, and Habit showed limited or marginal effects, with Social Influence revealing a significant but negative relationship. These findings suggest that internal psychological and motivational factors outweigh institutional support or habitual patterns in influencing the intention to adopt AI tools among pre-service teachers. The study concludes that strengthening personal efficacy, promoting perceived usefulness, and fostering intrinsic motivation through interactive exposure and purposeful integration of AI tools in teacher education programs will be essential for encouraging future use in classroom practice. Furthermore, this study imply that teacher education programs should focus on providing meaningful learning experiences that demonstrate the usefulness and enjoyability of AI tools. Structured, hands-on exposure to AI applications in classroom-relevant tasks may strengthen confidence, reduce anxiety, and promote exploratory engagement. Since constructs related to support, habit, and cost did not significantly influence intention, institutions should go beyond infrastructure and emphasize the motivational and personal relevance of AI integration. In addition, the negative effect of social influence suggests that adoption should be driven by internal readiness rather than external pressure. Programs should prioritize personalized training, self-directed engagement, and the development of instructional alignment to support sustainable and voluntary use of AI tools by future educators.

## References

[1] Abeer S Almogren, Waleed Mugahed Al-Rahmi, and Nisar Ahmed Dahri. 2024. Exploring factors influencing the acceptance of ChatGPT in higher education: A smart education perspective. *Heliyon* 10, 11 (2024), e31887. https://doi.org/https://doi.org/10.1016/j.heliyon.2024.e31887

[2] Amillia Amid and Rosseni Din. 2020. MOOCs Acceptance and Use in Higher Education Institutions through UTAUT2: An Overview. *Int. J. Acad. Res. Progress. Educ. Dev.* 9, 2 (2020), 559–569. https://doi.org/10.6007/IJARPED/v9-i2/7840

[3] Amillia Amid and Rosseni Din. 2024. A Review Studies on Acceptance and Use of MOOCs in HEIs: Applying UTAUT2. *Int. J. Acad. Res. Progress. Educ. Dev.* 13, 3 (2024), 5034–5049. https://doi.org/http://dx.doi.org/10.6007/IJARPED/v13-i3/17511

[4] Mekuriaw Genanew Asratie, Bantalem Derseh Wale, and Yibeltal Tadele Aylet. 2023. Effects of using educational technology tools to enhance EFL students' speaking performance. *Educ. Inf. Technol.* 28, 8 (2023), 10031–10051. https://doi.org/10.1007/s10639-022-11562-y

[5] Alberto Cattaneo, Maria-Luisa Schmitz, Philipp Gonon, Chiara Antonietti, Tessa Consoli, and Dominik Petko. 2025. The role of personal and contextual factors

when investigating technology integration in general and vocational education. *Comput. Human Behav.* 163, (2025), 108475. https://doi.org/https://doi.org/10.1016/j.chb.2024.108475
[6] Sara Dempster and Leah Peterson. 2023. Teacher Education Candidates Providing Educational Technology Professional Development to the University Community through Service-learning. *J. Serv. High. Educ.* 17, (2023), 1–16.
[7] Abid Haleem, Mohd Javaid, Mohd Asim Qadri, and Rajiv Suman. 2022. Understanding the role of digital technologies in education: A review. *Sustain. Oper. Comput.* 3, (2022), 275–285. https://doi.org/https://doi.org/10.1016/j.susoc.2022.05.004
[8] Insook Han, Shin Won Sug, and Yujung and Ko. 2017. The effect of student teaching experience and teacher beliefs on pre-service teachers' self-efficacy and intention to use technology in teaching. *Teach. Teach.* 23, 7 (October 2017), 829–842. https://doi.org/10.1080/13540602.2017.1322057
[9] Mosharrof Hosen, Samuel Ogbeibu, Beena Giridharan, Tat-Huei Cham, Weng Marc Lim, and Justin Paul. 2021. Individual motivation and social media influence on student knowledge sharing and learning performance: Evidence from an emerging economy. *Comput. Educ.* 172, (2021), 104262. https://doi.org/https://doi.org/10.1016/j.compedu.2021.104262
[10] Blal Idrees, Hugues Sampasa-Kanyinga, Hayley A Hamilton, and Jean-Philippe Chaput. 2024. Associations between problem technology use, life stress, and self-esteem among high school students. *BMC Public Health* 24, 1 (February 2024), 492. https://doi.org/10.1186/s12889-024-17963-7
[11] Davit Marikyan, Savvas, and Papagiannidis. 2023. Unified Theory of Acceptance and Use of Technology: A review. In *TheoryHub Book*, S. Papagiannidis (ed.). TheoryHub, 1–16.
[12] Afzal Sayed Munna and Abul Kalam. 2021. Teaching and learning process to enhance teaching effectiveness: a literature review. *Int. J. Humanit. Innov.* 4, 1 (2021), 1–4. Retrieved from https://files.eric.ed.gov/fulltext/ED610428.pdf
[13] Lergia I. Olivo and Kaitlin C. Alexander. 2021. When Left to Their Own Devices: Exploring Teacher Preference for Digital Learning Tools. *Adv. Glob. Educ. Res.* 4, (2021), 1–6. Retrieved from https://digitalcommons.usf.edu/cgi/viewcontent.cgi?article$=$1060&context$=$m3publishing
[14] Caleb Or. 2023. Examining Unified Theory of Acceptance and Use of Technology 2 through Meta-analytic Structural Equation Modelling. *J. Appl. Learn. Teach.* 6, 2 (2023), 283–293. https://doi.org/https://doi.org/10.37074/jalt.2023.6.2.7
[15] Isaac H Smith and Maryam Kouchaki. 2021. Ethical Learning: The Workplace as a Moral Laboratory for Character Development. *Soc. Issues Policy Rev.* 15, 1 (January 2021), 277–322. https://doi.org/https://doi.org/10.1111/sipr.12073
[16] Michele D Todino. 2025. Educational Technologies. *Encyclopedia 5*. https://doi.org/10.3390/encyclopedia5010023
[17] Detlef Urhahne and Lisette Wijnia. 2023. Theories of Motivation in Education: an Integrative Framework. *Educ. Psychol. Rev.* 35, 2 (2023), 45. https://doi.org/10.1007/s10648-023-09767-9
[18] Kai Wang, Qianqian Ruan, Xiaoxuan Zhang, Chunhua Fu, and Boyuan Duan. 2024. Pre-Service Teachers' GenAI Anxiety, Technology Self-Efficacy, and TPACK: Their Structural Relations with Behavioral Intention to Design GenAI-Assisted Teaching. *Behav. Sci. (Basel, Switzerland)* 14, 5 (April 2024). https://doi.org/10.3390/bs14050373
[19] Ying Xie, Chao Wan, and Kai Kong. 2024. Factors influencing Chinese pre-service teachers' behavioral intention and use behavior to adopt VR training system: based on the UTAUT2 model. *Humanit. Soc. Sci. Commun.* 11, 1 (2024), 1300. https://doi.org/10.1057/s41599-024-03832-6
[20] Liangyong Xue, Abdullah Mat Rashid, and Sha Ouyang. 2024. The Unified Theory of Acceptance and Use of Technology (UTAUT) in Higher Education: A Systematic Review. *SAGE Open* 14, 1 (January 2024), 21582440241229570. https://doi.org/10.1177/21582440241229570
[21] Ke Zhang and Ayse Begum Aslan. 2021. AI technologies for education: Recent research & future directions. *Comput. Educ. Artif. Intell.* 2, (2021), 100025. https://doi.org/https://doi.org/10.1016/j.caeai.2021.100025